\documentclass[prl, twocolumn ,floatfix,showpacs]{revtex4}
\usepackage{graphicx}

\renewcommand{\vec}{\mathbf}
\newcommand{\Rb}{${}^{87}$Rb}

\begin{document}
\title{Measurement of the ac Stark shift with a guided matter-wave interferometer}

\author{B. Deissler}
\author{K. J. Hughes}
\author{J. H. T. Burke}
\author{C. A. Sackett}
\email{sackett@virginia.edu}

\affiliation{Department of Physics, University of Virginia, Charlottesville, VA 22904}
\date{\today}

\begin{abstract}
We demonstrate the effectiveness of a guided-wave Bose-Einstein condensate interferometer for practical measurements. Taking advantage of the large arm separations obtainable in our interferometer, the energy levels of the \Rb ${}$ atoms in one arm of the interferometer are shifted by a calibrated laser beam. The resulting phase shifts are used to determine the ac polarizability at a range of frequencies near and at the atomic resonance. The measured values are in good agreement with theoretical expectations. However, we observe a broadening of the transition near the resonance, an indication of collective light scattering effects. This nonlinearity may prove useful for the production and control of squeezed quantum states.
\end{abstract}


\maketitle

Guided matter-wave interferometers have great potential for use in a variety of applications, from the precise measurement of inertial and gravitational effects to probes of chemical interactions \cite{Berman,Bongs04}. While interference with matter waves in free space has yielded impressive results \cite{Gustavson97,Ekstrom95}, a considerable amount of space is needed for the atoms to fall under the influence of gravity or for an atomic beam to propagate. Using atoms confined in a guiding potential can solve this problem, and permits larger arm separations and enclosed areas using more flexible geometries such as circular rings \cite{gupta05,arnold06}. So far, guided-wave interferometers have been ``proof-of-principle'' experiments in which interference has been demonstrated, but not applied to a practical measurement \cite{Kreutzmann04, shin04, Schumm05,wu05a,wang05,garcia06,Jo07}. In this Letter, we report the use of a guided-wave interferometer to measure the dynamic polarizability of atomic \Rb. We achieve results accurate to a few percent, comparable to or better than that achieved with other methods \cite{Kadar92,Sterr92,Morinaga93}. We also observe a broadening of the resonance line which we attribute to collective light scattering effects \cite{Javanainen94, You96}.

When atoms interact with the electric field of a beam of light, the energy levels of the atoms shift due to the ac Stark effect by an amount $U = -\frac{1}{2} \alpha \langle E^2 \rangle$. This effect is relevant to precision measurements, where the magnitude of such shifts must be well known \cite{Wieman87,Hemmer89}, and measurements of the polarizability $\alpha$ are important for verifying \emph{ab initio} calculations of dipole matrix elements \cite{safronova04}. Gradients of the Stark potential lead to optical dipole forces, which are used to great advantage for cooling and trapping atoms \cite{grimm00}, atom optics \cite{Adams94}, as well as quantum information \cite{Staanum02}. 

Combining the expression for the energy shift with the expression for the intensity of a beam of light, $I = c \epsilon_0 \langle E^2 \rangle$ gives
\begin{equation}
U = - \frac{1}{2 c \epsilon_0} \alpha I.
\end{equation}
Our experiment uses \Rb, which has two principal transitions from the $5S_{1/2}$ ground state, to the $5P_{1/2}$ state at 795~nm, and to the $5P_{3/2}$ state at 780~nm. Second-order perturbation theory gives the polarizability of an atomic state $|i\rangle$ as \cite{CohenQM}
\begin{equation}
\alpha_i(\omega_\ell) = \frac{2}{\hbar} \sum_{f\ne i} \frac{\omega_f}{\omega_f^2 - \omega_\ell^2}|\mu_{if}|^2,
\end{equation}
where $\omega_\ell$ is the frequency of the applied laser beam, $\omega_f$ is the transition frequency to an excited state $|f\rangle$, and $\mu_{if}$ is the transition dipole matrix element $\langle i |e\vec{r}| f \rangle$. This matrix element can be expressed in terms of the decay rate $\Gamma_{fi}$ of the excited state using
\begin{equation}
\Gamma_{fi} = \frac{\omega_f^3}{3\pi\epsilon_0\hbar c^3} |\mu_{if}|^2.
\end{equation}
We use this expression to normalize the dipole moments to the $5P_{3/2} \quad (F=3,m_F=3) \rightarrow 5S_{1/2} \quad (F=2,m_F=2)$ cycling transition rate, where $\Gamma_{33 \rightarrow 22} = \Gamma = 2\pi \times 6.065$ MHz is the total decay rate for the $5P_{3/2}$ states \cite{Volz96}. Then
\begin{equation}
\alpha_i(\omega_\ell) = 4\pi \epsilon_0 \frac{3 \Gamma c^3}{2\omega_{33}^3} \sum_{f \ne i} \frac{\omega_f}{\omega_f^2 - \omega_\ell^2} \left| \frac{\mu_{if}}{\mu_{33}} \right|^2.
\end{equation}
The ratios of dipole moments can be computed exactly from standard angular momentum algebra.

\begin{figure}
\includegraphics[width = 3.2in]{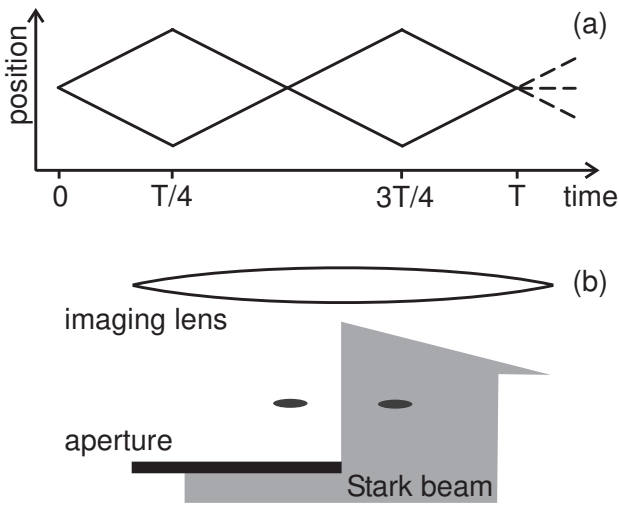}
\caption{(a) Trajectory of the atoms in the interferometer. The condensate is split at time $t=0$, reflected at times $T/4$ and $3T/4$, and recombined at time $T$, all with an off-resonant standing wave laser beam. The atoms are returned to rest by the recombination pulse with a probability that depends on the packets' differential phase. (b) Schematic of the experimental setup. The waveguide axis is in the horizontal direction. The Stark beam is apertured such that it only interacts with one packet of atoms at the maximum separation in the interferometer. This beam is imaged with the same camera that we use to observe the atoms.\label{fig:Starkassembly}}
\end{figure}
Our apparatus has been described in detail before \cite{garcia06,reeves05}. Briefly, we have a Bose-Einstein condensate of $3 \times 10^4$ \Rb ${}$ atoms in a harmonic waveguide generated by a time-orbiting potential (TOP) with transverse confinement frequencies of 3.3~Hz and 6~Hz and axial confinement of 1.1~Hz. An off-resonant standing-wave laser beam is used to evenly split the condensate with almost perfect efficiency into two packets which move in opposite directions along the waveguide axis with velocity $v_0 = 11.7$~mm/s \cite{hughes07}. The subsequent trajectories of the packets are shown in Fig.~\ref{fig:Starkassembly}(a). The packets move apart for a time $T/4$, where $T$ is the total time in the interferometer, and then the standing-wave laser beam is applied again to reflect the atoms, i.e. change their velocity $v_0 \leftrightarrow -v_0$. The atoms then propagate for a time $T/2$, after which another reflection pulse is applied. The atoms finally propagate for $T/4$ before the splitting pulse is applied again to recombine the packets. Two reflection operations are performed so that the two packets of atoms traverse nearly the same path and experience the same phase shifts from the guide potential \cite{Stickney07,burke07}. If the quantum state of the two clouds is unchanged during their time in the interferometer, the final recombination pulse is just the reverse action of the initial splitting pulse, and all the atoms return to rest. However, if the packets acquire a differential phase $\phi$, a fraction of the atoms will continue moving with speed $v_0$. After recombining, the atoms are allowed to separate for a short period of time before the distributions are imaged by absorption imaging. The ratio of the number of atoms at rest $N_0$ to the total number of atoms $N$ depends on the phase $\phi$,
\begin{equation}\label{eq:phase}
\frac{N_0}{N} = \frac{1}{2} \left( 1 + \cos \phi \right).
\end{equation}

For the experiments discussed here, the interferometer time is always $T = 40$ ms, for which we have nearly perfect visibility \cite{burke07}. The maximum separation of the two packets for this distance is $240$~$\mu$m, which is significantly larger than the cloud half-width of $55$~$\mu$m and allows for independent access to the two packets. This allows us to apply a laser beam (the ``Stark beam'') to one atom packet but not the other, as seen in Fig.~\ref{fig:Starkassembly}(b). 

In the experiment, the linearly polarized Stark beam is spatially filtered with a pinhole. An AOM is used to pulse on the Stark beam for a short time before and after the first reflect pulse, and a mechanical shutter is used to prevent any leakage light from the AOM from affecting the atoms. In addition, the power in the Stark beam is measured during the applied pulse with a photodiode and the intensity corrected accordingly to account for variations over time, which are generally less than 10\%. After completing the interferometer sequence and recombining the two clouds, the ratio $N_0/N$ is measured for different intensities and durations of the Stark beam. 

\begin{figure}
\includegraphics[width=3.2in]{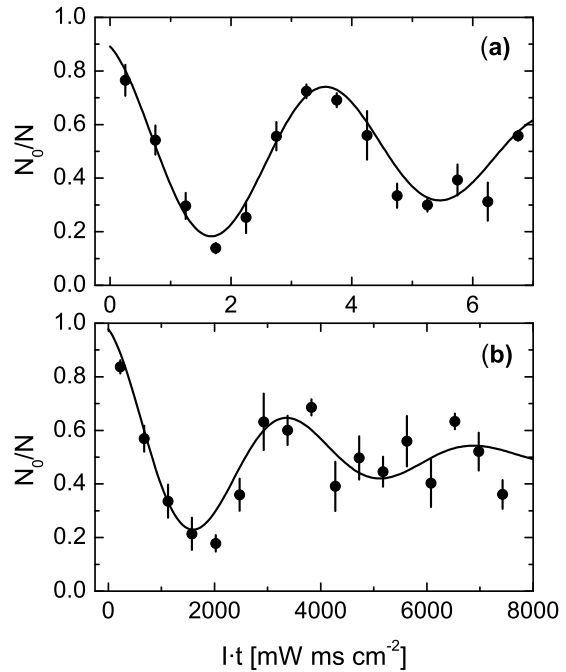}
\caption{Measurement of the phase shift from a laser beam at (a) 780.232~nm and (b) 808.37~nm. In both graphs, the solid line indicates a fit to the data. Since there is an increasing phase gradient across the atomic cloud with increasing phase, the visibility decays at higher intensities and longer times. In the graph, we have binned together groups of data with a similar product of measured intensity and pulse time. The bin size is $0.5$~mW ms cm${}^2$ in (a), $450$~mW ms cm${}^2$ in (b), and the error bars indicate the standard deviation of the mean within the bin.\label{fig:offresdata}}
\end{figure}
We performed this experiment at two different frequencies. For the first measurement, as seen in Fig.~\ref{fig:offresdata}(a), laser light locked to the $5S_{1/2}$, $F=1$ to $5P_{3/2}$, $F=2$ ``repump'' transition at 780.232~nm was used as the Stark beam. In two variations, a $475$~$\mu$s pulse was applied and the Stark beam intensity $I$ varied from $1$ to $15$~mW/cm${}^2$, or a beam with intensity $10.5$~mW/cm${}^2$ was applied for times $t$ ranging from $0$ to $675$~$\mu$s.  We fit the data to a function $f(It) = 1/2 + \exp(-\beta It) \cos\left(2\pi(It-x_0)/P\right)$, where $P$ is the period, $x_0$ is an overall phase offset, and $\beta$ is a decay constant reflecting the fact that intensity gradients in the beam induce spatial variations in the phase which eventually wash out the interference. We can extract the polarizabilities from the fit using $|\alpha| = 4\pi\epsilon_0 c\hbar/P$. For the data at 780.232~nm, the fit gives $|\alpha| = 4\pi\epsilon_0 \times (8.37 \pm 0.24) \times 10^{-25}$~m${}^3$, where the error is the statistical error from the fit. In order to get a theoretical value, we consider only transitions from the $5S_{1/2}$ ground state to the $5P_{3/2}$, $F=2,3$ excited states. The contributions from the Zeeman effect induced by the 20~G trap bias field and from $5P_{1/2}$ states are less than 0.1\%. Our calculations give a theoretical value of $|\alpha| = 4\pi\epsilon_0 \times 8.67 \times 10^{-25}$~m${}^3$, a deviation of $+3.5$\% relative to the experiment.

For the second measurement, seen in Fig.~\ref{fig:offresdata}(b), a free running laser diode at 808.37~nm was used. As in the previous case, a $1.276$~ms pulse with intensities ranging from $0$ to $6.6$~W/cm${}^2$ was applied to the atoms, or a beam with intensity $4.6$~W/cm${}^2$ was applied for various times from $0$ to $1.376$~ms. For this data, the fit gives $|\alpha| = 4\pi\epsilon_0 \times (9.48 \pm 0.25) \times 10^{-28}$~m${}^3$. Here, the Stark beam was focused down further, leading to higher intensities but also higher intensity gradients across the cloud. This explains the faster decay in the data at 808~nm than at 780~nm. A theoretical value was calculated using contributions from transitions to the $6P$ as well as the $5P$ states. Here, contributions from the hyperfine structure of the excited states and from higher $P$-states are less than 0.01\%, and are neglected. The calculated value for the polarizability is then $|\alpha| = (4\pi\epsilon_0) \times 9.14 \times 10^{-28}$~m${}^3$, differing from the measured value by $-3.7$\%. In both cases, we attribute the deviation between experiment and theory to inaccuracy of the intensity calibration.

To determine the intensity at the location of the atoms, we image the Stark beam using the same camera used to observe the atoms. We calibrate the camera by comparing the sum of the pixel values in an image of the unapertured Stark beam to the total power as measured by an optical power meter. Alternatively, the picture of the unapertured beam can be fit to a 2-dimensional Gaussian distribution to find its center and width. The intensity at the position of the atoms is then estimated as the value of the Gaussian function. This method effectively averages over high-frequency spatial noise in the beam, much of which is introduced by the imaging optics and is thus not present on the atoms. The intensities obtained using the two methods differ with a standard deviation of 4\%, and we use their average for our data. The other main source of error is the calibration of the Ophir PD200 power meter, which has a specified accuracy of 5\% at 780~nm. This accuracy is consistent with the difference observed when we compared the PD200 to a Coherent Lasermate meter. We therefore estimate a net calibration uncertainty of 6.5\%. This is comparable to the 8\% error estimated for a technique of Ref.~\cite{Kadar92}, which is to our knowledge the most precise previous measurement.

\begin{figure}
\includegraphics[width=3.2in]{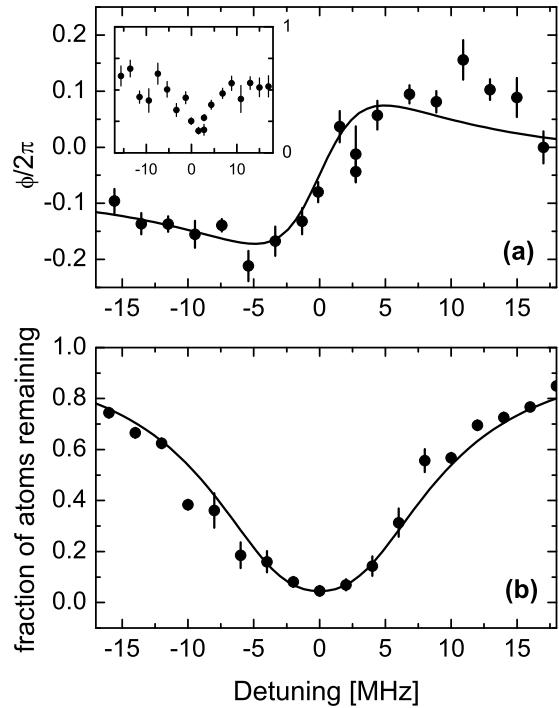}
\caption{(a) Measurement of the phase shift from a laser beam near the atomic resonance. The inset shows the visibility of the measured interferometer curve for various detunings of the Stark beam. (b) Loss of atoms due to the beam applied to one packet of atoms. The Rabi frequency and linewidth of the theory (solid line) were adjusted to match the data of the atom loss. These values were then used to calculate the theoretical curve in part (a).\label{fig:resdata}}
\end{figure}
Using a similar technique as the one described above, we used the interferometer to measure the ac Stark shift directly at an atomic resonance. The dispersion shape of the energy shift through resonance is well known, and can be measured using microwave spectroscopy of the ground state hyperfine transition \cite{Kastler63}. However, an atom interferometric technique would be necessary for atoms that lack ground state hyperfine structure. A measurement for a Bose-Einstein condensate is of particular interest due to the possibility for collective line broadening effects \cite{Javanainen94,You96}. 

In the resonant ac Stark shift measurement, we read out a phase shift caused by the Stark beam at a fixed time and intensity. By changing the phase $\theta$ of the standing wave during the recombination pulse, the fraction of atoms that remain at rest varies as $\cos^2(\phi/2 - 2\theta)$, where $\phi$ is the interferometer phase from Eq.~\ref{eq:phase}. We therefore take an interference curve by stepping through $\theta$ while keeping the Stark beam fixed. The phase shift of this curve gives $\phi$, which we measured for various detunings of the Stark beam. The results are shown in Fig.~\ref{fig:resdata}(a), with error bars from the fits of the interference curves. The overall offset of $-0.3$~rad is consistent with the residual phase induced by the guide itself with no Stark beam.

On resonance, photon scattering cannot be neglected. In our case, if an atom absorbs a photon, it leaves the condensate fraction and is lost from the magnetic trap. To avoid losing all the atoms, a relatively low intensity beam ($I \approx 40$~$\mu$W/cm${}^2$) is used that is applied for a short duration ($4$~$\mu$s) during the first reflect pulse. The small number of photons makes measuring the intensity with photodiodes and power meters difficult. Instead, we determine the intensity by measuring the atom loss when the experiment is performed without applying the recombination pulse. In this case, we observe just two packets, one of which has been exposed to the Stark beam and one that has not. From the ratio of the number of atoms in the packets we can accurately determine the atom loss, and thus the scattering rate. The results are shown in Fig.~\ref{fig:resdata}(b), along with a comparison to a calculation for a two-level system with an added decay term \cite{CohenAP}. We adjusted the Rabi frequency in the calculation to match the atom loss observed in the experiment. To obtain good agreement, however, we have found that a broadened linewidth of $\Gamma' = 2\pi \times (10.0 \pm 0.9)$~MHz is required, $1.6$ times larger than the natural linewidth $\Gamma$. The solid curve in Fig.~\ref{fig:resdata}(a) shows the phase shift calculated with these parameters. 

We confirmed the existence of this broadening by taking absorption lineshapes of a plain condensate, both in and out of the guide fields, and we attribute it to collective light scattering. Line broadening from such effects was predicted previously \cite{Javanainen94,You96}, and
superradiant scattering of an off-resonant beam has been observed in several experiments \cite{Inouye99}. The effect comes from Bose stimulation of the scattering process by the presence of recoiling atoms in the scattering channels. Javanainen has estimated the
linewidth of a condensate as
\begin{equation}
\Gamma' = \frac{3}{2} N \left( \frac{\lambda}{2 \pi L} \right)^2 \Gamma
\end{equation}
for a condensate of size L. If we use $L = (L_x L_y L_z)^{1/3} \approx 22$~$\mu$m for our Thomas-Fermi widths $L_i$, we obtain $\Gamma' = 1.5 \Gamma$, in reasonable agreement with the observations. The effect should be present only when there is a significant probability for each atom to scatter a photon, which explains how our measurements at large detuning are consistent with the unbroadened linewidth.

The presence of collective line broadening has important implications. It means that the phase shift imposed by a near-resonant Stark beam depends on $N$, which thus provides a source of controllable non-linearity for the quantum state of the atoms. For instance, condensate interferometers are expected to be limited by the fact that the atoms interact: the number of atoms in a packet has unavoidable quantum fluctuations, so the interaction
energy is uncertain \cite{Javanainen97}. This in turn imparts an uncertain phase shift that limits the coherence time of the interferometer \cite{Jo07}. However, if a near-resonant Stark beam can provide a number-dependent phase shift, this could be used to correct for the unknown interaction
phase and thus recover the coherence time. This would be beneficial even if it required the loss of some atoms due to the light scattering. More generally, any source of nonlinearity can produce
squeezed quantum states, which are useful for applications in quantum information \cite{Loudon}.

In conclusion, we have used a guided-wave atom interferometer to make practical high-quality measurements. The ac Stark shift for detuned light was measured with greater accuracy than possible with previous techniques. Measurements near resonance exhibit a broadened linewidth, which we are currently investigating in more detail. We also hope to extend our measurements to the dc polarizability by inserting one of the packets in between a pair of well-calibrated electric field plates. Through experiments such as these, guided-wave interferometers are beginning to make important contributions to metrology.

We thank T. F. Gallagher for the loan of equipment used in the experiments and R. R. Jones for helpful comments on the manuscript. This work was sponsored by the National Science Foundation (contract number PHY-0653349).


\end{document}